\documentclass[aps,prb,twocolumn,superscriptaddress,showpacs]{revtex4}
\usepackage{graphicx,color}
\begin{document}
\title{\bf Magnetic ordering of weakly coupled frustrated quantum spin chains}

\author{A.~A.~Zvyagin}
\affiliation{Max-Planck-Institut f\"ur Physik komplexer Systeme,
N\"othnitzer Str., 38, D-01187, Dresden, Germany}
\affiliation{B.~I.~Verkin Institute for Low Temperature Physics and
Engineering of the National Academy of Sciences of Ukraine, Lenin Ave., 47,
Kharkov, 61103, Ukraine}
\author{S.-L.~Drechsler}
\affiliation{Leibniz Institut f\"ur Festk\"orper- und Werkstoffforschung
Dresden, PF 270116 D-01171, Dresden, Germany}

\date{\today}

\begin{abstract}
The ordering temperature of a quasi-one-dimensional system,
consisting of weakly interacting quantum spin-1/2 chains with
antiferromagnetic spin-frustrating couplings (or zig-zag ladder)
is calculated. The results show that a quantum critical point
between two phases of the one-dimensional subsystem plays a
crucial role. If the one-dimensional subsystem is in the
antiferromagnetic-like phase in the ground state, similar to the
phase of a spin chain without frustration, weak couplings yield
magnetic ordering of the N\'eel type. For intra-chain
spin-frustrating interactions larger than the critical one (at
which the quantum phase transition takes place), the
quasi-one-dimensional spin system manifests a spiral magnetic
incommensurate ordering. The obtained results of our quantum
theory are compared with the quasi-classical approximations. The
calculated features of magnetic ordering are expected to be
generic for weakly coupled quantum spin chains with gapless
excitations and spin-frustrating nearest and next-nearest neighbor
interactions.
\end{abstract}

\maketitle

The interest in quasi-one-dimensional (quasi-1D) quantum spin
systems has grown considerably during the last decades. The
characteristic feature of quasi-1D magnets is the strong spin-spin
interaction along one space direction, much stronger than the
couplings along all other directions. The interest is motivated,
on the one hand, by the progress in preparation of substances with
well defined 1D subsystems. Another reason for studying quasi-1D
spin systems is the possibility to compare experimental data with
results of non-perturbative theories for 1D models. Also, such
systems often manifest quantum phase transitions that take place
in the ground state, and which are governed by other parameters
than the temperature, like external magnetic field, pressure,
concentration of impurities (internal pressure), etc. From the
experimental viewpoint, quasi-1D spin-1/2 systems differ from
other magnets due to special features in the behavior of their
characteristics. For example, the temperature dependence of the
magnetic susceptibility and the specific heat of quasi-1D spin
systems with dominant nearest neighbor (NN) interactions reveal
maxima (in a small external magnetic field at temperatures of the
order of the exchange coupling constant along the distinguished
direction). \cite{Zb} For spin systems, which 1D subsystems have a
gapless spectrum of low-lying excitations at temperatures, much
lower than the temperature of the maximum of the $T$-dependence of
the susceptibility, the latter and the specific heat often
manifest peculiarities, characteristic for phase transitions to
low-temperature magnetically ordered states at critical
temperatures.

Also, a great attention has been recently given to spin systems
with a spin frustration. For most of antiferromagnetic (AF)
systems the ground state corresponds to N\'eel-like
configurations. The standard quasi-classical description of AF
systems uses quantization of small deviations of vectors of
magnetizations (magnetic order parameters) of AF sublattices from
their steady-state configuration. However, such a description of
AF systems can be used for bipartite magnetic structures. For AF
systems with a spin frustration the competing interactions produce
a very high degeneracy of such steady-state configurations.
Therefore, in most cases it is hopeless to use the approximation
of magnetic sublattices. From the theoretical viewpoint the
situation becomes even worse in quasi-1D spin systems with a spin
frustration. For those systems quantum fluctuations are enhanced
due to peculiarities in the 1D density of states. This is why,
according to the famous Mermin-Wagner theorem, \cite{MW} 1D spin
systems with short-range interactions cannot have any magnetic
ordering even at $T=0$. Thus, approximate methods of theoretical
physics often produce significant errors in the description of
such systems. Hence, it is necessary to study them
non-perturbatively, better exactly, which is, fortunately,
possible for few cases for 1D quantum spin systems. Probably one
of the simplest and most known examples of a quantum spin system
with a spin frustration is the Heisenberg spin-$1/2$ chain with AF
nearest neighbor (NN) and AF next-nearest-neighbor (NNN)
interactions. The Hamiltonian of such a model can be written as
\begin{equation}
{\cal H}_{NNN} = J_1\sum_n {\bf S}_n{\bf S}_{n+1} +
J_2\sum_n {\bf S}_n{\bf S}_{n+2} \ ,
\label{Hnnn}
\end{equation}
where $J_1$ and $J_2$ are the couplings between NN and NNN,
respectively (here we consider only the case with an even number
of spins $N$). Such a system is equivalent to a zig-zag spin
ladder with obvious re-notation of indices. The system with the
Hamiltonian (\ref{Hnnn}) is, obviously, spin-frustrated. Several
limiting cases are known exactly. Namely, for $J_2=0$ (or for
$J_1=0$) the Hamiltonian (\ref{Hnnn}) is reduced to the
Hamiltonian of one (or two decoupled) Heisenberg AF spin-$1/2$
chain(s). The ground state for those cases is a non-degenerate
singlet, without long-range orderings, and the low-energy
excitations are gapless spinons. \cite{Zb} The other limiting
case, for which the ground state is known exactly, is the
so-called Majumdar-Ghosh point, $J_1=2J_2$. \cite{MG} In that case
the Hamiltonian of the zig-zag spin ladder can be re-written as
${\cal H}_{NNN} = (J_1/4)\sum_n ({\bf S}_n +{\bf S}_{n+1} + {\bf
S}_{n+2})^2 - 9N/4$. The ground state is given by two degenerate
singlets of the resonance valence bond type, without long-range
ordering, and the low-lying excitations are gapped. For other
values of the coupling constants it is, unfortunately, impossible,
to obtain exact answers. Nevertheless, an approximate bosonization
description and numerical calculations suggest that there is no
long-range magnetic ordering in the system, and that for $J_2 >
0.2411...J_1$ a spin gap is opened for the low-lying excitations.
\cite{ON} A quasi-classical approximation of the model yields the
following. If one replaces the spin operators by classical
vectors, two steady-state configurations are possible. The first
one is the period 2 commensurate and collinear AF N\'eel
configuration, stable for $J_1 > 4J_2$. The second one, stable for
$J_1 < 4J_2$, is a noncollinear incommensurate spiral magnetic
structure with the pitch angle $\cos \phi = -J_1/4J_2$. Such a
description implies a long-range magnetic order. This means that
it might be approximately valid for, e.g., a quasi-1D spin system,
consisting of weakly coupled 1D spin chains with NN and NNN AF
interactions, at temperatures, lower than the ordering
temperature. However, for quasi-1D spin systems with a spin gap
$\Delta_{sp}$ for low-energy excitations of their 1D subsystems, a
weak inter-chain coupling, as a rule, does not produce a magnetic
ordering. \cite{remark} (This is plausible at least for spin
systems with isotropic exchange interactions: the exponential
decay of the long-range spin-spin correlation function $\propto
\exp (-\xi/n)$ with a finite coherence length $\xi=\hbar
v/\pi\Delta_{sp}$, where $v$ is the Fermi velocity of the
low-lying spin excitations, conflicts with the magnetic order
requiring asymptotically nondecaying spin-spin correlation
functions.)

 On the other hand, it is clear that spin frustration in a
1D subsystem has to yield features in transitions to possible
magnetically ordered state for a quasi-1D system. Moreover, as
follows from Ref.~\onlinecite{expfr} (see also
Ref.~\onlinecite{inc}) despite the fact that for most of studied
compounds exchange constants satisfy the condition $J_2 >
0.2411...J_1$, the spin gap was not confirmed experimentally. To
describe such experimental situations (i.e.\ quasi-1D AF spin
systems with spin frustration of intra-chain interactions without
\textcolor{red}{a} spin gap and with a weak inter-chain coupling),
we consider another model, the Hamiltonian of which consists of
${\cal H}_{NNN}$ with multi-spin interaction. Such a model is
known to have gapless low-energy excitations. Those multi-spin
interactions do not change the spin frustration property from the
classical viewpoint. \cite{mult} The advantage of the proposed
model is the exact integrability: The model permits an exact
solution by means of the Bethe's ansatz. We do not state here,
naturally, that the model describes all features of the materials
of current experimental interest. \cite{expfr} However, many
properties of the model are similar to what was observed in
Ref.~\onlinecite{expfr}. At least, for this model the low-lying
excitations are gapless. Hence, from this viewpoint, they
qualitatively agree with the data of experiments, unlike the model
with the Hamiltonian ${\cal H}_{NNN}$. Multiple spin exchange
interactions are often present in oxides of transition metals,
where a direct exchange between magnetic ions is complimented by a
superexchange between magnetic ions via nonmagnetic ones.
\cite{exp} Models with multi-spin interactions are believed to be
closer to real quasi-1D magnets compared to standard ones with
only NN spin couplings. \cite{exp} Multi-spin exchange models were
introduced by Thouless already in 1965. \cite{Tho} Later similar
models were used to study some cuprates \cite{cupr} and spin
ladders. \cite{ladd} For the consistent explanation of several
experiments \cite{exp1} by means of inelastic neutron scattering,
optical conductivity and nuclear magnetic resonance one needs to
account for relatively large values of NNN spin-spin interactions
and multi-spin interactions between four neighboring sites of the
spin ladder (the so-called ring exchange). Similar four-spin
interactions were used recently in the theory of 2D quantum spin
systems, where they regulate the quantum phase transition between
the N\'eel-like ground state and the resonance valence bond solid
one. \cite{Sach} The Hamiltonian of the 1D subsystem of the
quasi-1D model, studied in our work, has the form
\begin{eqnarray}
&&{\cal H}_{1D} = {\cal H}_{NNN} + J_4\sum_n
\bigl( ({\bf S}_{n-1}{\bf S}_{n+1}) ({\bf S}_{n}{\bf S}_{n+2})
\nonumber \\
&&-  ({\bf S}_{n-1}{\bf S}_{n+2}) ({\bf S}_{n}{\bf S}_{n+1}) \bigr) \ .
\label{H2}
\end{eqnarray}
The model is also spin-frustrated. The classical counterpart of
the model (if one replaces the spin operators by classical
vectors) reveals a long-range magnetic ordering with a N\'eel
steady-state configuration, or with a spiral magnetic structure,
where the four-spin ring exchange renormalizes the spiral pitch
angle as $\cos \phi = -2J_1/(8J_2 -J_4)$. However, quantum
properties of the model with the Hamiltonian ${\cal H}_{1D}$
differ from the one with ${\cal H}_{NNN}$ in a much more drastic
way than of their classical counterparts. This can be seen from
the exact solution (the exactly solvable model was introduced in
Ref.~\onlinecite{MT}), which is known for the parametrization of
coupling constants $J_1=J(1-x)$, $J_2 =Jx/2$, $J_4=2Jx$ for any
$J$ and $x$ (in what follows we shall consider $J>0$, $x >0$). For
$x=0$ the model describes the Heisenberg spin-$1/2$ AF chain. As
one can see from exact results, the high degeneracy of low-energy
states, caused by the spin frustration of NN and NNN interactions,
is removed by adding the ring exchange, which is also
spin-frustrated. According to the exactly known properties, the
ground state of the model Eq.~(\ref{H2}) depends on the values of
the coupling constant $x$ and an external magnetic field $H$.
\cite{obz} For large values of $H$ the model is in the
spin-saturated (ferromagnetic) phase. This phase has a trivial
long-range magnetic order and gapped low-lying excitations. It is
divided from other phases by the line of the second order quantum
phase transition. For low values of $x$ and $H$ the model is in
the phase, which behavior is similar in properties to the phase of
the Heisenberg spin-$1/2$ AF chain in a weak magnetic field.
\cite{Zb} Notice that the model is in this phase for $x < x_{cr} =
4/\pi^2$ even at $H=0$. At $x_{cr}$ a second order quantum phase
transition takes place. For $x > x_{cr}$ and $H=0$ the model
reveals an incommensurate ordering with nonzero spontaneous
magnetization. For nonzero values of $h$ and large enough values
of $x$ the model is in the incommensurate magnetic phase with
nonzero weak magnetization. \cite{obz} The degeneracy in the
direction of the spontaneous magnetization can be removed, if one
first puts the system into an external magnetic field, and, then,
removes the field. These two phases are divided from each other by
the line of the second order quantum phase transition (quantum
critical point).
\begin{figure}
\begin{center}
\includegraphics[width=0.5\textwidth]{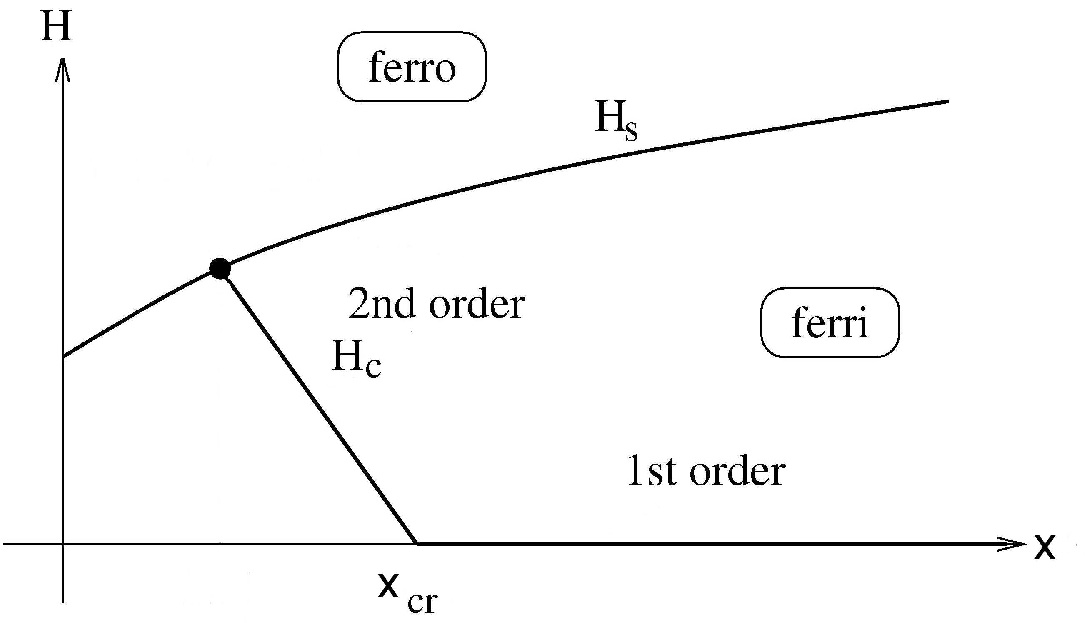}
\end{center}
\caption{The phase diagram $H-x$ of the one-dimensional integrable
spin model. At the lines $H_s$ and $H_c$ second order quantum
phase transitions to the ferromagnetic (spin-saturated) phase, and
the ferrimagnetic spiral one, respectively, take place. In the
point $H=0$, $x=x_{cr}$ the second order quantum phase transition
takes place. At the line $H=0$ for $x> x_{cr}$ the first order
phase transition takes place.} \label{fig1}
\end{figure}
Fig.~1 shows the ground state phase diagram of the one-dimensional
integrable model. The mentioned quantum phase transitions can be
observed in the temperature behavior of thermodynamic
characteristics of the model, like the magnetic susceptibility and
the specific heat, that were also calculated exactly. \cite{ZK}

The goal of our present study is to find how the weak coupling
between frustrated spin chains can produce magnetic orderings, and
what are specific features of such orderings.

According to the above, one can suppose two different types of
low-temperature magnetic ordering in the quasi-1D system under
consideration. For the first one, the N\'eel ordering, one can
write the magnetization of the $n$-th site, e.g., as ${\bf M}_n =
M{\bf e}_z + (-1)^n m_N {\bf e}_x$, where ${\bf e}_{x,z}$ are the
unit vectors in the $x$- or $z$ directions, $M$ is the average
magnetization, and $m_N$ is the staggered magnetization in the
direction, perpendicular to the external field. Another
possibility is the 3D generalization of the ground state
incommensurate phase of the 1D subsystem, the spiral
incommensurate state with the magnetization of the $n$-th site,
written as ${\bf M}_n = M{\bf e}_z + \cos (Qn) m_{sp} {\bf e}_z$,
where $m_{sp}$ is the modulated component of the $z$-projection of
the magnetization around the average magnetization $M$, and $Q =
\pi(1-2M)$ is the wave vector of the 1D modulated structure.

As usually, we study the weak inter-chain coupling $J'$ in the mean field
approximation. In that approximation in the N\'eel phase we can write the
mean field Hamiltonian of the 1D subsystem as
\begin{eqnarray}
&&{\cal H}^{mf}_N = {\cal H}_{1D} - (H -zJ'M)\sum_n S^z_n
\nonumber \\
&&- h_N\sum_n (-1)^n S_n^x + {\rm const} \ ,
\label{HN}
\end{eqnarray}
where $h_N =zJ'm_N$, and $z$ is the coordination number. For the spiral phase
the mean field Hamiltonian is
\begin{eqnarray}
&&{\cal H}^{mf}_{sp} = {\cal H}_{1D} - (H -zJ'M)\sum_n S^z_n
\nonumber \\
&&- h_{sp}\sum_n\cos (Qn)S_n^z + {\rm const} \ ,
\label{Hsp}
\end{eqnarray}
where $h_{sp} =zJ'm_{sp}$. Renormalization group-like approach
\cite{Zb} implies that both $h_N$ and $h_{sp}$ are relevant
perturbations. They generate spin gaps $\Delta E_N \sim
h_N^{2/(4-\eta)}$ and $\Delta E_{sp} \sim h_{sp}^{2\eta/(4\eta
-1)}$, respectively, for low-energy excitations. Here $\eta$ is
the correlation function exponent, see below, which determines the
asymptotical behavior of the spin-spin correlation functions of
the 1D subsystem in the conformal limit. \cite{ZK}

The order parameters $m_N$ and $m_{sp}$ (or $h_N$ and $h_{sp}$)
have to be determined self-consistently. In the mean field
approximation the corresponding self-consistency equations can be
written as
\begin{equation}
m_{N,sp} = M_{N,sp}(H,h_{N,sp},T) \ , \label{selfcons}
\end{equation}
where $M_{N,sp}(H,h_{N,sp},T)$ is the magnetization per site of
the 1D subsystem in the effective field $H - MzJ'$ and $h_{N,sp}$
at the temperature $T$. Then the transition temperature to the
ordered state has to be determined from the equation
\begin{eqnarray}
&&1 = zJ'\chi_{N,st} \ ,
\nonumber \\
&&\chi_{N,st} = \biggl( {\partial
M_{N,st}(h,h_{N,sp},T)\over \partial h_{N,sp}} \biggr)_{h_{N,sp} \to 0} \ .
\label{Tord}
\end{eqnarray}
The non-uniform susceptibilities of the 1D subsystem at low
temperature can be found as
\begin{equation}
\chi_{\alpha}(q,T) = -i \sum_n \int dt e^{-iqn} \Theta (t)\langle
[S^{\alpha}(n,t), S^{\alpha}(0,0)]\rangle_{T} \ ,
\label{sus}
\end{equation}
where $q$ is the wave vector, $\alpha =x,z$, and $\langle ... \rangle_T$
denotes the thermal average at the temperature $T$. Asymptotics of the
correlation functions for an integrable spin chain can be obtained in the
conformal field theory limit. \cite{Zb} For the model with the Hamiltonian
${\cal H}_{1D}$ it was done in Ref.~\onlinecite{ZK}, and we can write for the
ground state correlation functions
\begin{eqnarray}
&&\langle S^z_n S^z_0 \rangle \approx M^2 +{B^*\cos(Qn) \over
[n^2-(vt)^2]^{\theta_z}} + \dots \ ,
\nonumber \\
&&\langle S^x_n S^x_0 \rangle \approx (-1)^n {C \over
[n^2-(vt)^2]^{\theta_{\perp}}} +  \dots \ ,
\label{corr}
\end{eqnarray}
where $v$ is the Fermi velocity of low-energy excitations,
$\theta_z = Z^2 (\equiv 1/2\eta)$, $\theta_{\perp} = 1/4Z^2 \equiv
(\eta/2)$, $Z$ is the dressed charge of low-lying excitations,
$B^*$, and $C$ are nonuniversal constants. In particular, we see,
that the symmetry of the ground state is lower than the symmetry
of the Hamiltonian, caused by the ordering, i.e. for our model one
deals with the manifestation of the Goldstone theorem.
Eqs.~(\ref{corr}) can be extended for weak nonzero temperatures
using the conformal mapping $(n-vt) \to (2v/\pi T)\sinh [\pi
T(n-vt)]$. \cite{Zb} Then, calculating susceptibilities according
to Eq.~(\ref{sus}) (we use the main approximation) for $q=\pi$ for
the N\'eel phase and for $q=Q$ for the spiral incommensurate
phase, we obtain the expressions for the ordering temperatures
\begin{equation}
T_N ={v\over 2\pi} \biggl[ C {zJ'\over v}\sin \left({\pi \eta\over 2}\right)
B^2\biggl({\eta\over 4},
{2-\eta\over 2}\biggr)\biggr]^{1\over 2-\eta} \ ,
\label{TN}
\end{equation}
and
\begin{equation}
T_{sp} ={v\over 2\pi} \biggl[B^* {zJ'\over v}\sin\left({\pi\over
2\eta} \right) B^2\biggl({1\over 4\eta}, {2\eta-1\over
2\eta}\biggr) \biggr]^{\eta\over 2\eta -1} \ , \label{Tsp}
\end{equation}
where $B(x,y)$ is the Euler's beta function. In those expressions the Fermi
velocity and the critical exponent $\eta$ (or the dressed charge $Z$) can be
calculated using the Bethe ansatz results. \cite{ZK} Then, the question to be
answered is: Which ordering temperature, $T_N$ or $T_{sp}$, is higher for the
quasi-1D spin chain with the spin frustration?

In what follows we limit ourselves with the case $H=0$ for
simplicity. In this situation the effective Fermi velocity can be
written as $v = (\pi/2)J|1- (x/x_{cr})|$. \cite{ZK} Consider first
the ground state phase $x < x_{cr}$ for a 1D subsystem, which is
similar to the ground state of the Heisenberg spin-1/2 chain with
only NN AF interactions. In this case we have $M=0$ and $\eta =1$.
\cite{ZK} For this case we can use $B^*=C \sim 0.2$. \cite{L} One
can see that in this case (i.e. $\eta =1$) $T_N =T_{sp} =
(CzJ'/2\pi)B^2(1/4,1/2)$. We see that the critical temperature
does not depend on $J$ and $x$ (obviously, any nonzero magnetic
field $h \ne 0$, or an inclusion of a magnetic anisotropy will
change this result). To get the $J$- and $x$-dependences even for
$H=0$ and for the magnetically isotropic case, one has to include
logarithmic corrections, \cite{Zb} reproducing the known result
for $x=0$, \cite{S}
\begin{equation}
T_N = {CzJ'\over 2\pi}B^2(1/4,1/2)\sqrt{\ln\left( {\pi^2
J|1-(x/x_{cr})| \over CzJ'B^2(1/4,1/2)  }\right) } \ , \label{TN1}
\end{equation}
which is valid, naturally, when the argument of the logarithm is
larger or equal to 1. Next, let us consider the ground state of
the 1D subsystem for $x > x_{cr}$, which ground state has a
spontaneous magnetic ordering. This spontaneous magnetization
$M(x>x_{cr}) \ne 0$ is connected with holes in the ground state
distribution of quantum numbers, called rapidities, which form the
Dirac sea of the 1D subsystem. \cite{MT,ZK} Those holes appear
only for $x > x_{cr}$. \cite{MT,obz,ZK} Notice that in the
previous case, $x < x_{cr}$, there are no holes in the Dirac sea,
and the ground state rapidities can have any value in the range
$-\infty,\dots, \infty$. \cite{MT,obz,ZK} It is impossible to find
an analytic solution for $\eta$ in this case. We see that $\eta =1
+a$, with $0 \le a \le a_{max} < 1$ when $1 \le x/x_{cr} <
\infty$. Unfortunately, in this case we cannot obtain the values
of the non-universal constants. We can only suppose that they are
also of the order of $0.05-0.2$. \cite{L} It is easy to see that
for most of the values of $J$, $J'$ and $x$ the temperature of the
transition to the spiral incommensurate state for $x > x_{cr}$ is
higher than the N\'eel temperature.
\begin{figure}
\begin{center}
\includegraphics[width=0.35\textwidth]{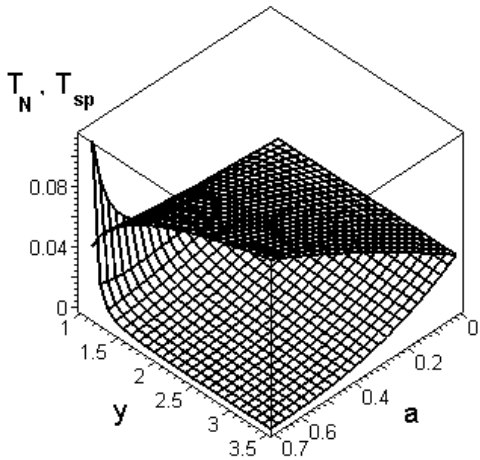}
\end{center}
\caption{The N\'eel temperature (the lower surface) and the
ordering temperature of the transition to the spiral
incommensurate phase for a quasi-1D spin-1/2 chain with the spin
frustration, caused by AF NN, NNN interactions and the ring
exchange as functions of the deviation of the critical exponent
$a=1-\eta$ and $y=x/x_{cr}$, which shows how close the quantum
critical point $y=1$ is.} \label{fig2}
\end{figure}
In Fig.~2 we plotted $T_N$ (lower surface) and $T_{sp}$ (upper
surface), for $J=1$, $J'= 0.01$, $z=4$ and $B=C=0.2$ as functions
of $a$ and $y=(x/x_{cr})$. It turns out that $a$ is a function of
$x/x_{cr}$ also, but, unfortunately, one cannot find this
dependence analytically. Notice that for $a =0$ ($\eta =1$,
$x=x_{cr}$) both critical temperatures coincide. The N\'eel
temperature can be larger than the temperature of the transition
to the spiral incommensurate phase only in the vicinity of the
quantum phase transition $x = x_{cr}$ for $\eta$ being very large
(close to 2, which seems to be an over-estimation, cf.
Ref~\onlinecite{MT}). For all other values of $x > x_{cr}$ we get
$T_{sp} > T_N$. Hence, we can conclude that for a quasi-1D system,
consisting of weakly coupled spin-1/2 chains with AF
spin-frustrating NN and NNN interactions and with the four-spin
ring exchange, the low-temperature ordering depends on the
behavior of 1D subsystems. For small values of the NNN
interaction, the quasi-1D system undergoes a transition to the
magnetically ordered AF N\'eel state. On the other hand, if the
exchange constant of the NNN interactions exceeds the critical
value, at which a quantum phase transition to the incommensurate
state with the weak spontaneous magnetization takes place, a weak
coupling between 1D subsystems produces the transition to the
magnetically ordered incommensurate spiral state. The ordering
temperature in the latter is
\begin{eqnarray}
&&T_{sp} ={J|1-y|\over 4} \biggl[ B^* {2zJ'\over \pi J|1-y|}
\sin\left({\pi \over 2(1+a)}\right) \times
\nonumber \\
&&B^2\biggl({1\over 4(1+a)}, {1+2a\over 2(1+a)}\biggr)
\biggr]^{1+a\over 1+2a} \ .
\label{Tsp1}
\end{eqnarray}
Our quantum analysis qualitatively agrees with the quasi-classical
description of the considered system. However, a difference
between the quantum system and its quasi-classical counterpart
exists: In the quantum system magnetic ordering takes place for
$J' \ne 0$, only. We expect analogous expressions to Eqs. (11,12)
to be valid also for the other critical point between a
ferromagnetic and a spiral phase, i.e.\ if $J_1 < 0$. In this
context, the determination of the magnetic order below their
low-temperature phase transitions at few K for Li$_2$ZrCuO$_4$ and
Pb[Cu(SO)$_4$(OH)$_2$], \cite{baran} both being close to that
critical point, would be of interest.

We expect that an inter-chain ferromagnetic interaction,
\cite{though} for $x < x_{cr}$ has to produce the AF
low-temperature ordering of a special type. Namely, we expect low
temperature ordered phase, which consists of ferromagnetic planes
with alternating magnetizations of planes. On the other hand, for
$x > x_{cr}$, a ferromagnetic inter-chain interaction is expected
to produce ferrimagnetic low-temperature ordering with a nonzero
spontaneous total magnetization.

Finally, let us consider what happens, if one studies the
situation with only NN and NNN couplings, without the ring
exchange. In that case, for $J_2 < 0.2411...J_1$ the quasi-1D
system undergoes a phase transition to the N\'eel state, due to
weak couplings between chains. For $J_2 > 0.2411...J_1$, a spin
gap is opened for low-lying excitations of the 1D subsystem, and
the ordering temperature goes to zero. This case seems to
contradict known experiments, in which magnetic ordering was
observed even for $J_2 > 0.2411...J_1$. \cite{expfr} Therefore, we
can conclude, that for some real compounds with the properties of
quasi-1D spin systems with spin-frustrating interactions in their
1D subsystems some additional spin-spin interactions, like the
ring exchange, studied in this paper, probably exist, which close
the spin gap and give rise to magnetic orderings at low
temperature.

In summary, the ordering temperature of a quasi-one-dimensional
system, consisting of weakly interacting quantum spin-1/2 chains
with antiferromagnetic spin-frustrating couplings (or zig-zag spin
ladder) is calculated. Our results show that the quantum critical
point between the two phases of the 1D-subsystem plays an
important role. If the one-dimensional subsystem is in the ground
state in an antiferromagnetic-like phase, similar to the phase of
a spin chain without frustration, weak couplings yield a magnetic
ordering of the N\'eel type. On the other hand, for intra-chain
spin-frustrating interactions larger than the critical one (at
which the quantum phase transition takes place), an incommensurate
spiral magnetic ordering of the quasi-one-dimensional spin system
takes place. The obtained results of the quantum theory are
compared with the quasi-classical approximations. We expect that
the calculated features of the magnetic ordering are generic for
weakly coupled quantum spin chains with gapless excitations and
with spin-frustrating nearest and next-nearest neighbor
interactions. While up to now we do not know quasi-one-dimensional
systems with NN AF spin interactions and large AF NNN ones in the
spiral phase at low temperatures (cf. Ref.~\onlinecite{expfr},
see, though, Refs.~\onlinecite{Sr,Ti}), we believe that our
results can be used for comparison with the observed temperatures
of magnetic orderings in other spin frustrated quasi-1D quantum
spin systems.

\section*{Acknowledgement}

The Deutsche Forschungsgemeinschaft (S.-L.\ D.) is acknowledged
for financial support. We thank J. Richter, R. Kuzian, and H.
Rosner for interest and discussions.


\begin{thebibliography}{99}
\bibitem{Zb} See, e.g., A.~A.~Zvyagin {\em Finite Size Effects in Correlated
Electron Models: Exact Results}, Imperial College Press, London, 2005.
\bibitem{MW} N.~D.~Mermin and H.~Wagner, Phys. Rev. Lett. {\bf 17}, 1133
(1966).
\bibitem{MG} C.~K.~Majumdar and D.~K.~Ghosh, J. Math. Phys. {\bf 10}, 1388
(1969).
\bibitem{ON} K.~Okamoto and K.~Nomura, Phys. Lett. A {\bf 169}, 433 (1992).
\bibitem{remark} Here we ignore the possibility of a spin-Peierls transition
related to distant dependent intra-chain exchange integrals and a
soft enough lattice. In this case (corresponding roughly speaking
to the condensation of singlets on short bonds of a dimerized
chain) no local magnetization occurs below the phase transition
and in that sense there is also no magnetic ordering. The
dimerization is strongly supported by $J_2<0.7J_1$.
\bibitem{expfr} M.~Matsuda and K.~Katsumata, J. Magn. Magn. Mater. {\bf
140-145}, 1671 (1995); H.~Kikuchi, H.~Hagasawa, Y.~Ajiro,
T.~Asano, and T.~Goto, Physica B {\bf 284-288}, 1631 (2000);
N.~Maeshima, M.~Hagiwara, Y.~Narumi, K.~Kindo, T.~C.~Kobayasi, and
K.~Okunishi, J. Phys.: Condensed Matter {\bf 15}, 3607 (2003);
M.~Hase, K.~Ozawa, and N.~Shinya, Phys. Rev. B {\bf 68}, 214421
(2003).
\bibitem{inc} T.~Masuda, A.~Zheludev, A.~Bush, M.~Markina, and A.~Vasiliev,
Phys. Rev. Lett. {\bf 94}, 039706 (2005); S.-L.~Drechsler, J.~M\'alek,
J.~Richter, A.~S.~Moskvin, A.~A.~Gippius, and H.~Rosner, Phys. Rev. Lett.
{\bf 94}, 039705 (2005); A.~A.~Gippius, E.~N.~Morozova, A.~S.~Moskvin,
A.~V.~Zalessky, A.~A.~Bush, M.~Baenitz, H.~Rosner, and S.-L.~Drechsler,
Phys. Rev. B {\bf 70}, 020406(R) (2004); H.-A.~Krug von Nidda, L.~E.~Svistov,
M.~V.~Eremin, R.~M.~Eremina, A.~Loidl, V.~Kataev, A.~Validov, A.~Prokofiev,
and W.~A\ss mus, Phys. Rev. B {\bf 65}, 134445 (2002).
\bibitem{mult} See, e.g., G.~Misguich, B.~Bernu, C.~Lhuillier, and
C.~Waldtmann, Phys. Rev. Lett. {\bf 81}, 1098 (1998).
\bibitem{exp} A.~Zheludev, M.~Kenzelmann, S.~Raymond, E.~Ressouche, T.~Masuda,
K.~Kakurai, S.~Maslov, I.~Tsukada, K.~Uchinokura, and A.~Wildes,
Phys. Rev. Lett. {\bf 85}, 4799 (2000); I.~Tsukada, J.~Takeya, T.~Masuda, and
K.~Uchinokura, Phys. Rev. B {\bf 62}, R6061 (2000); M.~Kohgi, K.~Iwasa,
J.~M.~Mignot, B.~Fak, P.~Gegenwart, M.~Lang, A.~Ochiai, H.~Aoki, and
T.~Suzuki, Phys. Rev. Lett. {\bf 86}, 2439 (2001).
\bibitem{Tho} D.~J.~Thouless, Proc. Phys. Soc. {\bf 86}, 893 (1965).
\bibitem{cupr} Y.~Honda, Y.~Kuramoto, and T.~Watanabe, Phys. Rev. B
{\bf 47}, 11329 (1993); H.~J.~Schmidt, and Y.~Kuramoto, Physica C {\bf
167}, 263 (1990).
\bibitem{ladd} S.~Brehmer, H.-J.~Mikeska, M.~M\"uller, N.~Nagaosa, and
S.~Uchida, Phys. Rev. B {\bf 60}, 329 (1999); M.~Matsuda,
K.~Katsumara, R.~S.~Eccleston, S.~Brehmer, and H.-J.~Mikeska,
J. Appl. Phys. {\bf 87}, 6271 (2000).
\bibitem{exp1} T.~Imai, K.~R.~Thurber, K.~M.~Shen, A.~W.~Hunt, and
F.~C.~Chou, Phys. Rev. Lett. {\bf 81}, 220 (1998); R.~S.~Eccleston,
M.~Uehara, J.~Akimitsu, H.~Eisaki, N.~Motoyama, and S.~I.~Uchida, {\em
ibid.} {\bf 81}, 1702 (1998); M.~Windt, M.~Gr\"uninger, T.~Nunner,
C.~Knetter, K.~P.~Schmidt, G.~S.~Uhrig, T.~Kopp, A.~Freimuth,
U.~Ammerahl, B.~B\"uchner, and A.~Revcolevschi, {\em ibid.} {\bf 87},
127002 (2001); K.~Magishi, S.~Matsumoto, Y.~Kitaoka, K.~Ishida,
K.~Asayama, M.~Uehara, T.~Nagata, and J.~Akimitsu, Phys. Rev. B {\bf
57}, 11533 (1998).
\bibitem{Sach} See, e.g., S.~Sachdev, Nature Physics  {\bf 4}, 173
(2008) (preprint arXiv:cond-mat/0711.3015); A.~W.~Sandvik, Phys.
Rev. Lett. {\bf 98}, 227202 (2007).
\bibitem{MT} N.~Muramoto and M.~Takahashi, J. Phys. Soc. Jpn. {\bf 68}, 2098
(1999).
\bibitem{obz} For the review, use A.~A.~Zvyagin, J. Phys. A {\bf 34}, R21
(2001).
\bibitem{ZK} A.~A.~Zvyagin and A.~Kl\"umper, Phys. Rev. B {\bf 68}, 144426
(2003).
\bibitem{L} S.~Lukyanov and A.~Zamolodchikov, Nucl. Phys. B {\bf 493}, 571
(1997); S.~Lukyanov, Phys. Rev. B {\bf 59}, 11163 (1999); S.~Lukyanov and
V.~Terras, Nucl. Phys. B {\bf 654}, 323 (2003).
\bibitem{S} H.~J.~Schulz, Phys. Rev. Lett. {\bf 77}, 2790 (1996); I.~Affleck
and M.~Oshikawa, Phys. Rev. B {\bf 60}, 1038 (1999).
\bibitem{baran} M.\ Baran,
V.A.\ Jedrzejczak, H.\ Szymczak, V.\ Maltsev, G.\ Kamieniarz, G.\
Szukowski, C.\ Loison, A.\ Ormeci, S.-L.\ Drechsler, and H.\
Rosner, Phys.\ Stat.\ Sol.\ c, {\bf 3}, 220 (2006).
\bibitem{though} As far as we know, this case is not yet realized among known
quasi-1D spin-ladder systems, except, probably, CuGeO$_3$, which,
however, has gapped low-energy excitations, and, hence, has no
long range AF ordering.
\bibitem{Sr} I.~A.~Zaliznyak, C.~Broholm, M.~Kibune, M.~Nohara, and H.~Takagi,
Phys. Rev.\ Lett.\ {\bf 83}, 5370 (1999).
\bibitem{Ti} G.~J.~Nilsen, H.~M.~R\/onnow, 
A.~M.~L\"auchli, F.~P.~A.~Fabbiani,
J.~Sanchez-Benitez, K.~V.~Kamenev, and A.~Harrison, Chem. Mater.
{\bf 20}, 8 (2008).
\end{thebibliography}
\end{document}